\documentstyle[11pt,paspconf,epsf]{article}

\begin{document}

\title{Dynamics in the cosmological mass function (or, why does the
Press \& Schechter work?)}

\author{Pierluigi Monaco}
\affil{Institute of Astronomy, Madingley Road, Cambridge CB3 0HA, UK}

\begin{abstract}
The Press \& Schechter ``numerical recipe'' is briefly reviewed,
together with the recently proposed dynamical mass function theory, in
which the mass function is constructed by using the powerful
Lagrangian perturbation theory.  The dynamical mass function is found
in good agreement with the recent N-body simulations of Governato et
al. (1998), in the case of an Einstein-de Sitter Universe.  The
definition of collapse, the relation between mass and smoothing
radius, and the definition of structure in 1D Universes are discussed.
A detailed comparison of the dynamical mass function to simulations
reveals that the orbit-crossed regions in the simulation are correctly
reproduced, while the fragmentation of the collapsed medium into
structures cannot be done in a univocal way.  Finally, we try to
answer the question: why the hell does the Press \& Schechter work?
\end{abstract}


\section{Introduction}

The mass distribution of collapsed dark matter clumps (called mass
function) is one of the most important pieces of information one wants
to obtain from a cosmological model.  These clumps are the sites where
the most important astrophysical events take place; they are
associated to proto-galactic halos, and to galaxy groups and clusters.

Unfortunately, non-linear gravitational collapse is an unsolved
problem in the general case, and the only way to get a fair estimate
of the mass function is through N-body simulations or through (semi-)
analytical approximations.

The first approximation to the cosmological mass function was
proposed\ldots by Doroshkevich (1967), and later by Press \& Schechter
(1974; hereafter PS).  The PS ``numerical recipe'' (as recently
referred to by Efstathiou) can be summarized as follows:

\begin{enumerate}
\item Take the initial density field, at a very early time, and smooth
it on the scale $R$, such that the variance of the field is $\Lambda
\equiv \sigma^2(R)$; we can't work with power at very small scales,
which becomes non-linear very early.
\item Follow the linear evolution of the density field: the density in
each point is rescaled by the same time-dependent function $b(t)$, the
linear growing mode.
\item If the linearly evolved density gets as large as a threshold
$\delta_c\sim 1$, then the density in those points has gone
non-linear!  In other terms, it has collapsed.
\item Compute the fraction of collapsed mass $\Omega(<\Lambda)$ as the
probability that the linearly evolved density is larger than the
threshold $\delta_c$.  Underdensities (1/2 of mass with Gaussian
statistics) in linear theory get more and more underdense, so no
wonder that only 50\% of mass is able to collapse; to achieve the
correct normalization, multiply the result by two, and don't ask why!
\item If the field is smoothed at a scale $R$, it is quite reasonable
to assume that a typical mass $M\sim\varrho_0 R^3$ will form
($\varrho_0$ is the comoving average mass density).  Then the mass
function $n(M)$ is computed through the following ``golden rule''
(as referred to by Cavaliere):

\begin{equation} Mn(M)dM \simeq \varrho_0 \frac{d\Omega}{d\Lambda}
\left|\frac{d\Lambda}{dM}\right| dM \label{eq:ps} \end{equation}

\end{enumerate}

Despite its weaknesses, the PS numerical recipe works much better than
expected.  Governato et al. (1998) give the most recent comparison of
the PS formula to large numerical simulations: the PS is a good
approximation of the N-body curves (found with different clump-finding
algorithms), for masses similar to or larger than the typical mass
$M_*(z)$ which forms at a given redshift.  The $\delta_c$ parameter is
not very different from the value 1.69 guessed by the spherical
collapse model, except that it tends to get as small as 1.5 at $z=1$
(for a standard CDM model).

The PS did not receive much attention until 1988, when the first
``large'' N-body simulations of Efstathiou et al. (1988) and Carlberg
\& Couchman (1989) showed a good agreement with it.  The mystery of
the ``fudge factor'' of 2 was solved by Peacock \& Heavens (1990) and
Bond et al. (1991), who approached the ``cloud-in-cloud'' problem in a
rigorous way (small structures can be included in larger collapsing
ones, even though their density is not larger than the threshold, and
taking this into account increases the total fraction of collapsed
mass to 1).  It was again a surprise to see that the PS formula,
including the factor of 2, is exactly recovered if the linear density
field is smoothed with a sharp filter in the $k$-space.

The history of the mass function theory is reviewed in Monaco (1998),
but it can be effectively summarized in a sentence: there is a simple,
effective and wrong way to describe the cosmological mass function.
Wrong of course does not refer to the results but to the whole
procedure.

\section{The mass function is an intrinsically Lagrangian quantity}

Once the ``statistical'' problem of achieving the correct
normalization is solved, the worst defect of the PS recipe is that it
completely neglects the complexities of gravitational dynamics, which
is treated just at the linear level.

The Lagrangian picture of fluid dynamics turns out to be an ideal
frame in which to develop a semi-analytical theory for the mass
function, able to go beyond linear theory and spherical collapse.
Before going on, it is appropriate to criticize the use of spherical
collapse in cosmology, for a number of reasons:

\begin{enumerate}
\item With spherical symmetry the collapse is clearly identified with
the formation of a singularity.  In the more general case, the
definition of collapse is quite tricky, as we are going to show in the
following.
\item Virialization is assumed to have taken place immediately after
collapse.  This neglects the many complicated transients which
dominate structures like galaxy clusters.
\item Further accretion is spherical.  In the real case accretion
comes preferentially from the filamentary network.
\end{enumerate}

The construction of a ``dynamical'' mass function is described in
detail in Monaco (1995; 1997a,b) and reviewed in Monaco (1998).  The
dynamics of gravitational collapse is described though the powerful
Lagrangian perturbation theory, truncated to third order (which is
necessary to describe collapse, as the second order truncation gives
wrong results in underdensities).  It is useful to recall that the
first term in the perturbation series is the Zel'dovich (1970)
approximation.  If the infinitesimal mass element is described as a
homogeneous ellipsoid, then ellipsoidal collapse turns out to be a
useful truncation of the Lagrangian series.

In the general case in which no symmetry is imposed to the initial
density field, the definition of collapse itself is tricky.  There is
a natural way to define collapse, and it is when orbits start to cross
each other (orbit crossing, hereafter OC).  The typical pancake
collapse takes place at OC; this means infinite density, infinite
shear, shock waves in a subdominant baryon component, and, last but
not least, no reliable solution after it!  Then, as long as structures
are searched for as high-density clumps, this definition makes sense;
moreover, it marks the onset of multi-stream, highly non-linear
dynamics and of shock-heating of gas.

However, it might be argued that OC is too generous in mixing ``real
clumps'', as galaxy clusters, with the filamentary network existent
outside clusters.  If the mass element is modeled as an ellipsoid,
then OC corresponds to the collapse on its first axis.  As proposed
for instance by Lee \& Shandarin (1997), one might extrapolate the
Lagrangian series so as to wait for the collapse of the element on the
third axis, and then get the mass function of fully virialized clumps.
The difference between the two collapse definitions (OC and third-axis
collapse) is well illustrated in the following example.  Consider a
spherical peak in a homogeneous Universe, the profile of which is not
a simple top-hat but decreases from the center.  While the central
mass element collapses in a spherical way, all the other mass elements
are subject to axial symmetry, so that their typical collapse is of
the spindle kind: two axes collapse, the third one does not collapse
at all. In the meantime each shell collapses in a perfectly spherical
way.  As a consequence, while the OC criterion selects all the
collapsed mass elements, just one mass element collapses on the third
axis.  Then, the total mass of the spherically collapsing clump,
according to the collapse definition based on third-axis collapse, is
embarrassingly vanishing.

This example highlights two important facts: (i) the local geometry of
collapse is different from the global one; (ii) third-axis collapse is
good for picking the seeds of collapsed structures, but not for
obtaining their total mass, while OC can get the correct
normalization, but mixes virialized clumps with filamentary
transients.  The second conclusion is confirmed by the fact that Lee
\& Shandarin (1997) need a fudge factor of 12.5 to achieve a good
normalization for their mass function.

\begin{figure}
\plotone{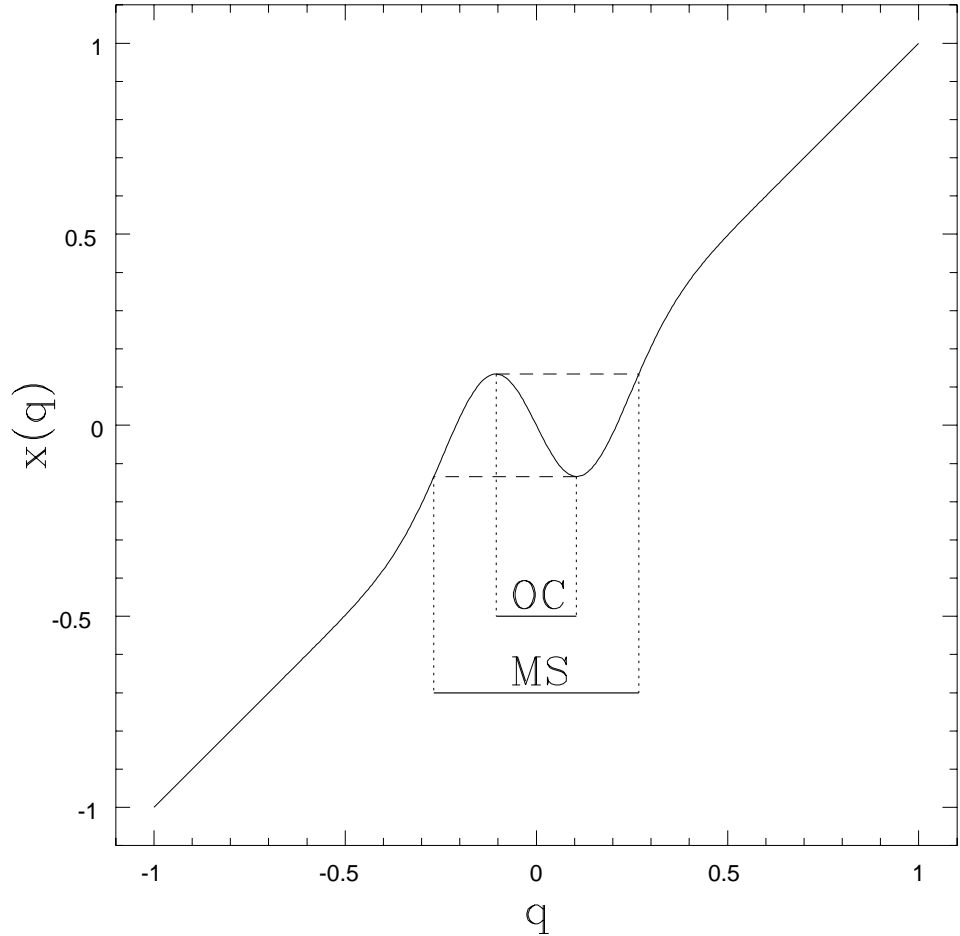}
\caption{Difference between orbit crossing and multi streaming.}
\end{figure}

Another limit of the OC definition is the following.  Fig. 1 shows a
1D sketch of a collapsing structure; $q$ is the initial position of a
particle (the {\it Lagrangian} coordinate) and $x$ is its final
position (the {\it Eulerian coordinate}).  The triple valued region in
the $x(q)$ relation constitutes the whole multi-stream region, but the
OC condition is able to select just the zones with negative slope in
the $x(q)$ relation.  In other words, the OC condition misses the mass
which is actually infalling on the structure, picking it only after
its first crossing of the structure.  This problem somehow compensates
for the excessive generosity of OC in finding the collapsed mass: it
just singles out matter which has had some time to relax!  Note also
that this problem is not present in the adhesion model, in which the
caustics have an infinitesimal width: it is connected to the finite
size of caustics.

Once collapse is defined, one can compute directly from initial
conditions the collapse time of each mass element.  It is convenient
to express time in terms of the linear growing mode $b(t)$, (to
factorize out the dependence on cosmology), and to define the
following quantity:

\begin{equation} F(q)=\frac{1}{b_c(q)} \end{equation}

where $b_c$ is the collapse ``time'' for the element $q$.  In the
linear theory case, $F=\delta/\delta_c$.  Note that, being $F$ (the
inverse of) a time, any threshold $F_c$ is just given by the inverse
time at which the mass function is wanted. There are no free
parameters in this theory.

The $F(q)$ function depends also on the smoothing radius used for the
smoothing.  To calculate the mass function with the correct
normalization, it is possible to extend the ``excursion set approach''
of Bond et al. (1991) to the case of the non-Gaussian $F$ process,
smoothed with a sharp $k$-space filter.  The case of Gaussian
smoothing is treated as in Peacock \& Heavens (1990).  Note that a
PS-like approach (Monaco 1995), with no fudge factor, is enough to
obtain a fair estimate of the mass function, as the fraction of points
which will never collapse is not 50\%, as in linear theory, but it is
smaller than 8\%.

The shape of the filter should be chosen so as to optimize the
dynamical predictions; the Gaussian filter is usually suggested.  It
is very interesting to see that the dynamical mass function gives
meaningful results only in the case of Gaussian smoothing.  In
particular, it is very similar to a PS curve with a $\delta_c$
parameter about 1.5, smaller than the spherical 1.69 value.  This is
explained by the fact that the spherical collapse neglects tidal
forces, which are able to speed up the collapse, thus forming larger
objects.

\begin{figure}
\plottwo{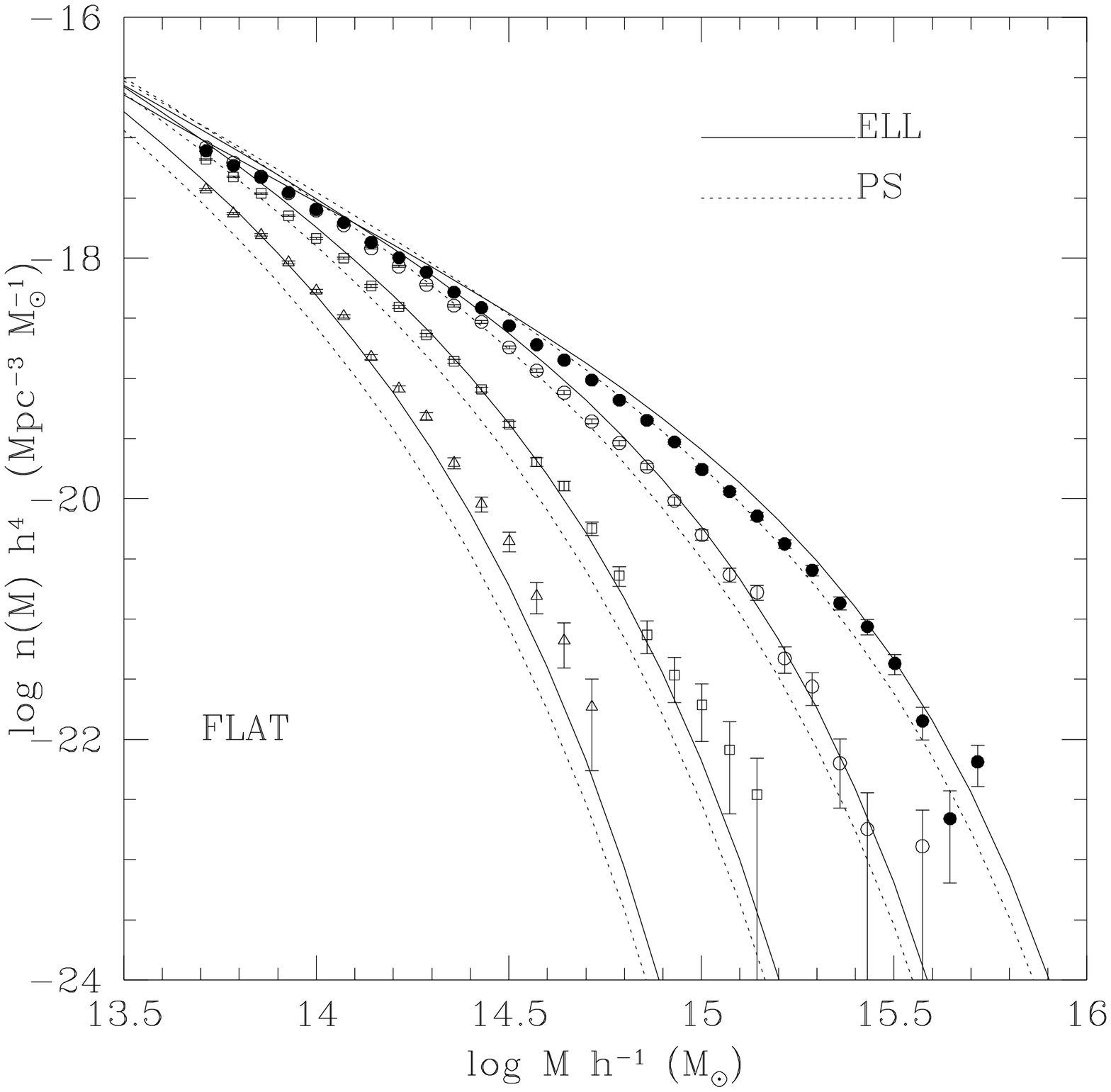}{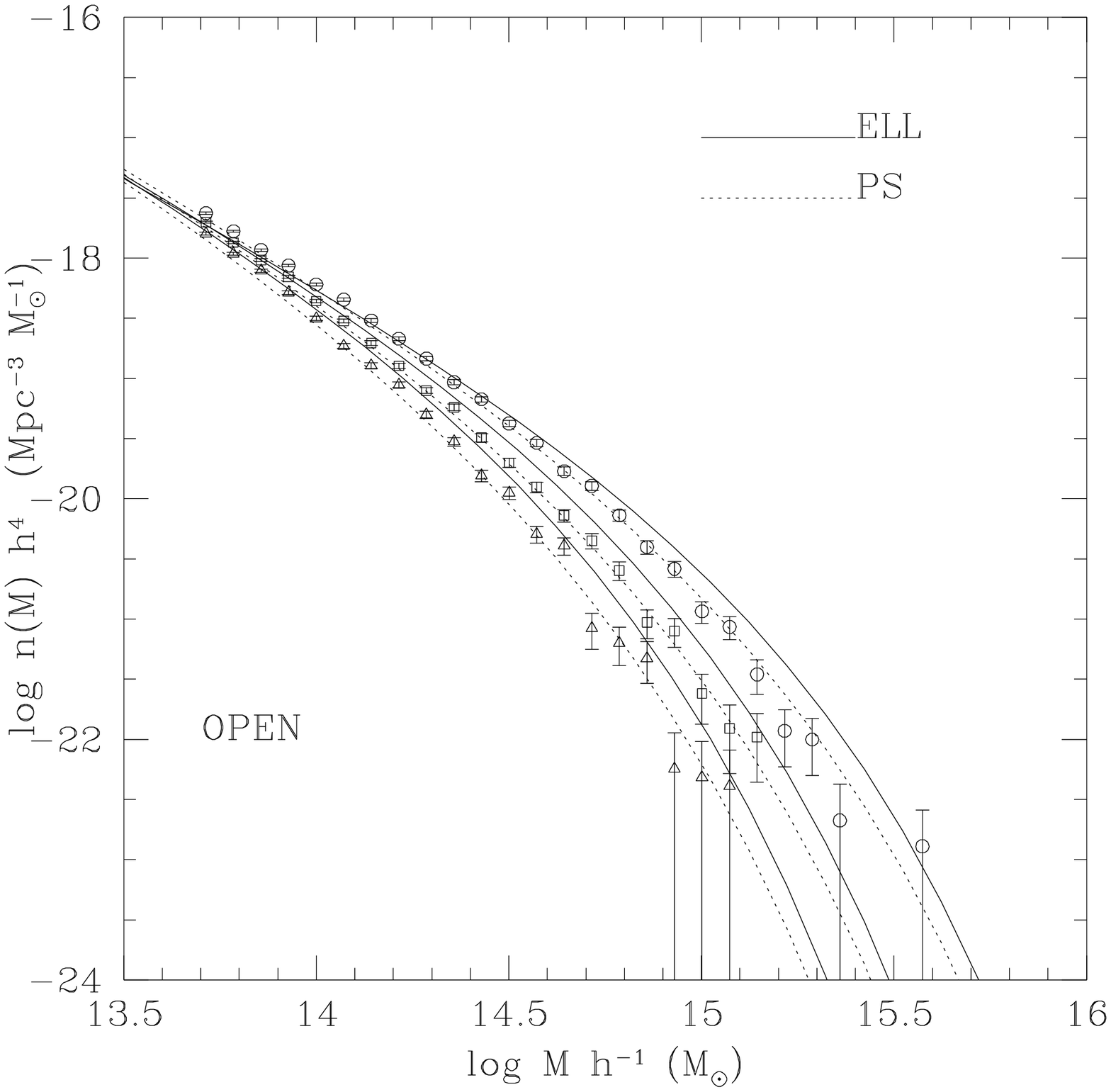}
\caption{The dynamical mass function compared to the N-body
simulations of Governato et al. (1998) (see text for details).}
\end{figure}

Fig. 2 shows a comparison of the dynamical mass function (with
Gaussian smoothing) with the simulations of Governato et al. (1998);
the left panel shows the results for a flat Einstein-de Sitter
Universe (standard CDM, $h=0.5$), at redshift $z=1.86$, 1.13, 0.43 and
0, while the right panel shows the results for an open Universe
($\Omega=0.3$, $h=0.75$, no cosmological constant), at redshift $z=1$,
0.58, and 0.  Groups are found with the standard friends-of-friends
algorithm, with linking-length 0.2 (corrected in the open case as
described in the paper); the authors have verified that the use of
other clump-finding algorithms does not change much the results.  The
N-body mass functions are compared with the standard PS prediction
(with $\delta_c=1.69$) and with the dynamical mass function, using
ellipsoidal collapse.  It must be noted that the 3rd-order prediction
is known to underestimate the number of large-mass objects (Monaco
1997a): in this range the ellipsoidal prediction performs better.

Despite the absence of free parameters to tune, the dynamical mass
function fits the N-body results remarkably well in the Einstein-de
Sitter case: the agreement is improved with respect to the PS formula,
with the possible exception of the $z=0$ mass function which is
slightly overestimated, while the $z=1.86$ mass function is
underestimated in the highest-mass tail.  The trend of the best-fit
$delta_c$ with redshift is not reproduced at this stage.  On the other
hand, the PS formula describes well the results of the open
simulation, better than the dynamical mass function.  Note also that
the sharp $k$-space predictions lie more than a factor two above the
Gaussian curves: at variance with the PS theory, the sharp $k$-space
filter does not seem a good choice in this context.

Another conclusion of the dynamical mass function theory is that the
small-mass tail cannot be predicted in a robust way.  Many factors
hamper a reliable prediction of small-mass clumps; most importantly,
the Lagrangian perturbation theory does not converge to a solution,
and any error in the definition of large masses strongly influences
the number of small-mass objects.

\section{From smoothing radius to mass: a 1D perspective}

Once the mass function theory is founded on a solid dynamical basis,
it is worth investigating in some detail the geometrical problem of
going beyond the ``golden rule'' (Eq.\ref{eq:ps}).  In other words, we
want to obtain the distribution of the masses $M$ of the objects which
form at a given smoothing radius $R$, so as to relate the fraction of
collapsed mass to the mass function in a rigorous way.

\begin{figure}
\plottwo{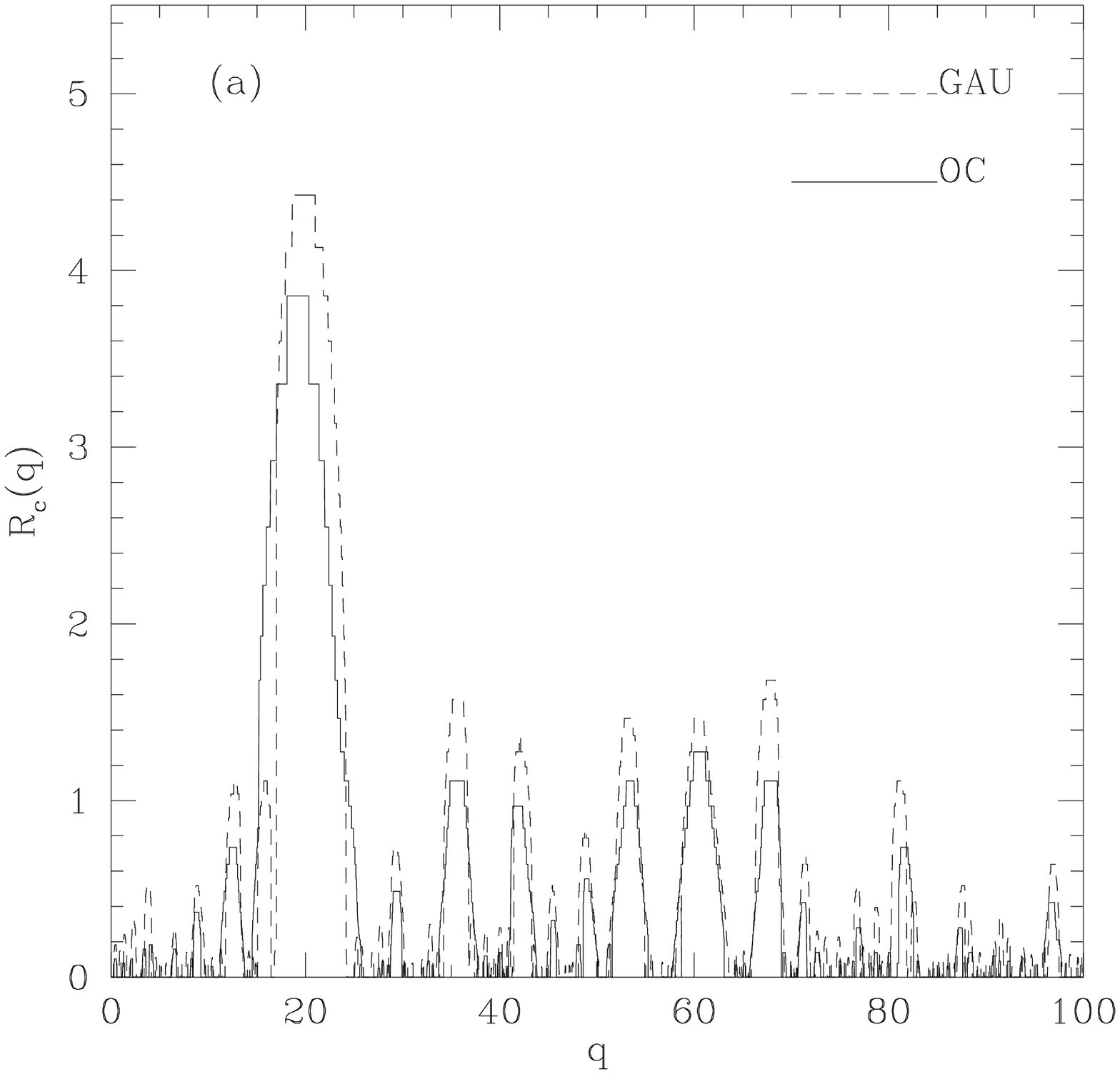}{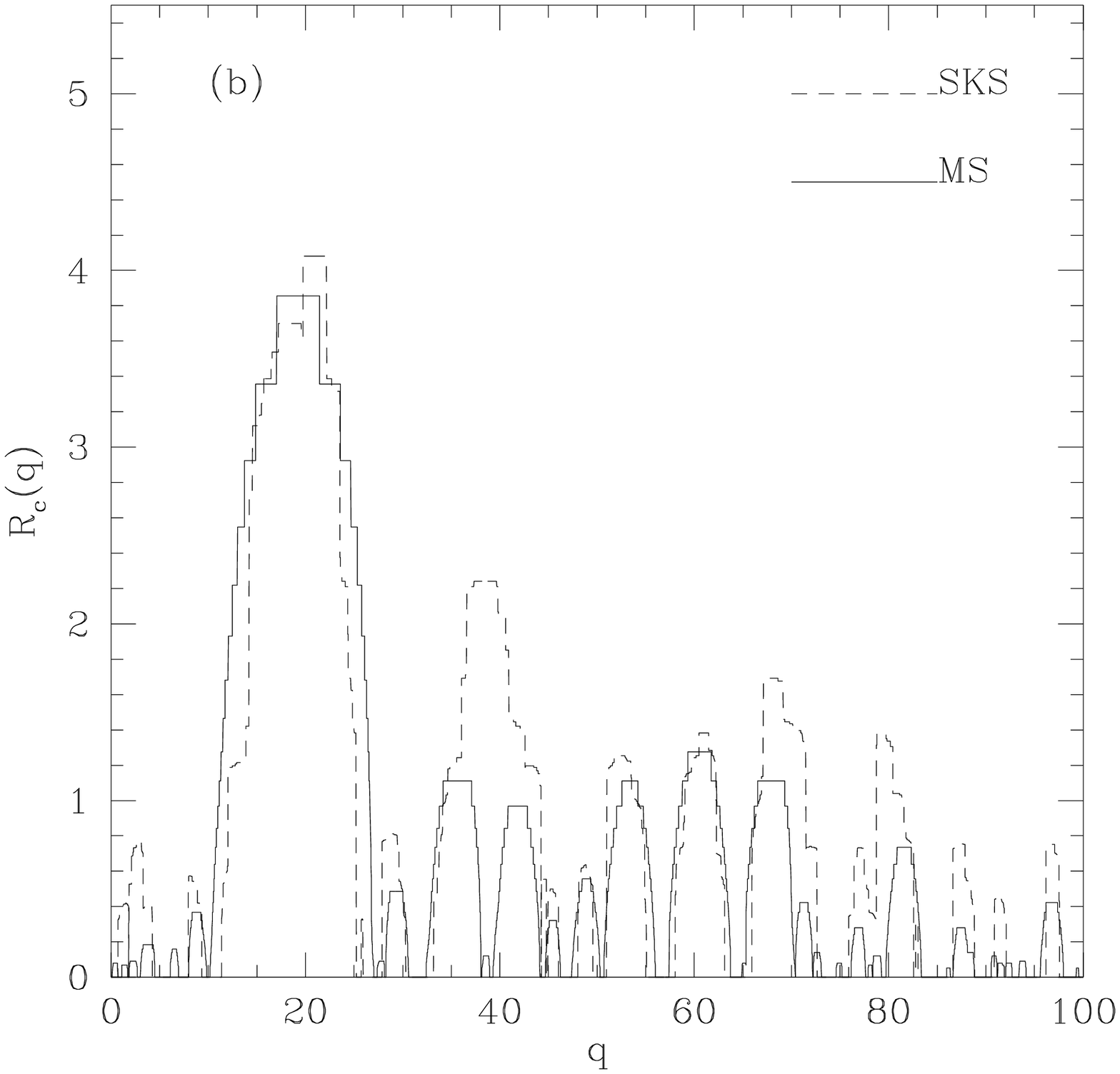}
\caption{$R_c(q)$ curves in a 1D Universe.}
\end{figure}

It is very useful to consider this problem first in 1D.  Monaco \&
Murante (1998) have solved the mass function problem in 1D in all its
details; this is a very instructing game, which gives many precious
hints on how to face the more complex 3D problem.  It is useful to
construct a function $R_c(q)$, which gives, for each point $q$ in the
Lagrangian space the largest smoothing radius $R_c$ at which the point
is predicted to collapse within a fixed time $t$.  Fig. 3 shows some
examples of these curve, in the case of a white noise power spectrum.
Both the predictions based on Gaussian and sharp $k$-space filters are
shown.  ``Objects'' can be easily associated to connected segments of
collapsed points; the objects present at a given radius $R$ are then
given by the intersection of the $R_c(q)$ curve with a line of
constant $R$.  Note also that, being $R_c$ the {\it largest} smoothing
radius at which collapse takes place, the collapsed regions are
already corrected for the cloud-in-cloud problem.

It is important to note that, when $R$ is decreased, the collapsed
objects show a typical behaviour: their mass (=length) starts growing,
and then saturates to a certain value, remaining approximately stable
for a significant range in $R$.  This stabilization is remarkable, and
necessary in order to define the mass of an object in any meaningful
way.

Under the hypothesis that the objects follow a well defined average
growing curve $G(R)$, it is possible to connect the fraction of collapsed
mass with the abundance of objects of mass $M$ in the following way:

\begin{equation} \Omega(>R) = \frac{1}{\varrho_0}\int_0^\infty
M n(M) G(R,M) dM \label{eq:deconv} \end{equation}

\begin{figure}
\plottwo{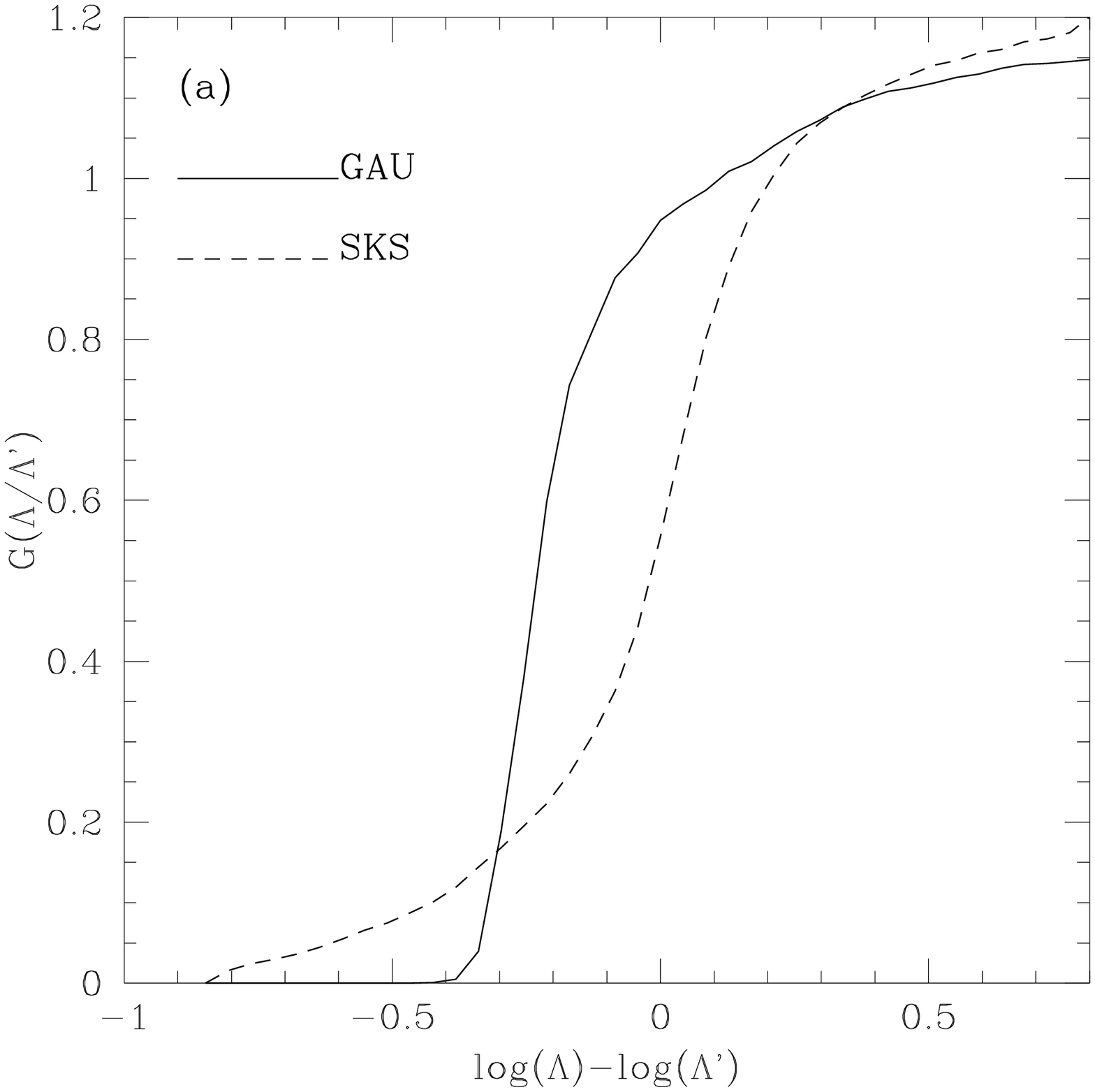}{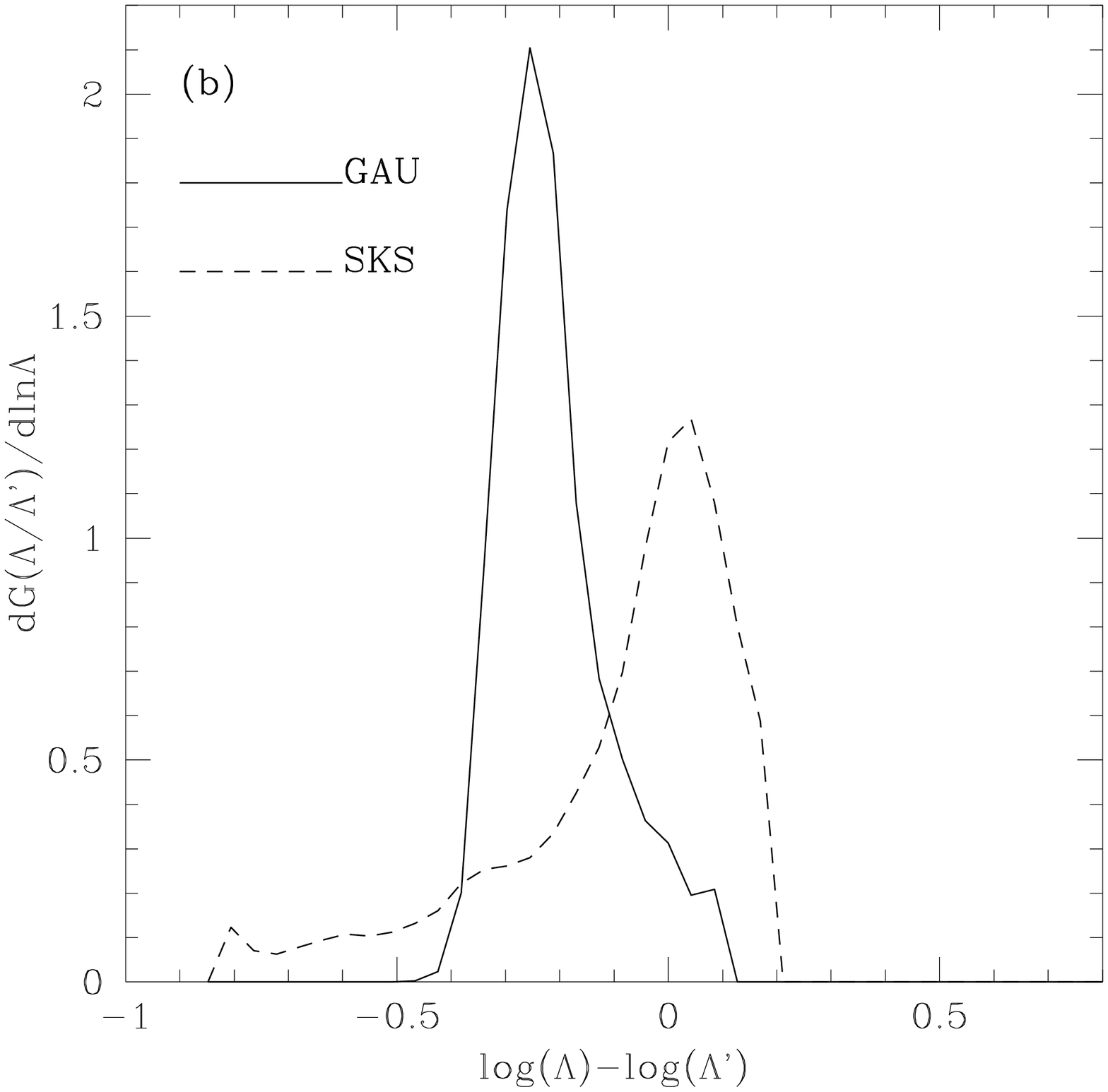}
\caption{Cumulative and differential growing curves.}
\end{figure}

With some algebra it is possible to write Eq.~(\ref{eq:deconv}) in the
form of a convolution of the mass function with the differential
growing curve.  Fig. 4 shows the cumulative and differential growing
curves in the Gaussian and sharp $k$-space cases (both $M$ and $R$ are
expressed in terms of the variance $\Lambda$); while the shar
$k$-space curves are not particularly nice, the Gaussian differential
curve shows one pronounced and narrow peak.  It is easy to see that
the golden rule corresponds to a Heavyside $\Theta$ function for the
growing curve, or a Dirac $\delta$ function for the differential one;
in other words, the golden rule corresponds to approximating the peaks
shown in Fig. 3 with many rectangles, the height ($R$) and width ($M$)
of which are related.  In the realistic case of a non-$\delta$
differential growing curve, the mass function must be deconvolved from
it.  However, for most practical purposes the mean value of the
differential growing curve is enough: the golden rule strikes back!
even though this average value could be different from that inferred
from the mass of the filter, and could depend on the power spectrum.

Such a dependence of the average mass on the power spectrum is
probably at the origin of the observed decrease of $\delta_c$ with
redshift in CDM Universes (Governato et al. 1998).  The slope of the
power spectrum, at the scale corresponding to the onset of
non-linearity, gets smaller (and negative) at higher redshift, which
means that the smoothed density field gets more correlated.  It is
then quite natural to expect larger masses, i.e. a decrease of
$\delta_c$, at higher redshift.

\section{OC regions in N-body simulations}

The 1D problem is also a good starting point for performing a detailed
comparison of the mass function theory with simulations.  Note also
that in 1D linear theory (with $\delta_c=1$) and Zel'dovich
approximation are equivalent, and moreover they are an exact solution
up to OC.  What is predicted by the theory is not the mass function of
``virialized'' structures (the word ``virialized'' is widely misused
in cosmology: we don't understand the physical processes which lead to
virialization, and then we don't know how to quantify them), nor is it
the mass function of halos found with the friends-of-friends or
whatever algorithm.  The predictions refer to regions which are in OC.
It is then useful to find OC regions directly in the simulations, and
to compare them to the theoretical predictions; in a second time, the
relation between such OC regions and groups, as found by some
clump-finding algorithm, can be investigated.

OC regions in the simulations are found by smoothing the $x(q)$ map
(the final positions of particles as a function of their initial
positions) in the Lagrangian space, then considering all the points
with negative slope as collapsed.  The OC is then scale-dependent; one
can construct the $R_c(q)$ curve also in this case, and compare it with
the predicted one.  From OC, it is possible to reconstruct the whole
multi-stream (MS) regions, as the multi-valued regions in the $x(q)$
relation.  The $R_c$ curves for the OC and MS regions are shown in
Fig. 3, together with the predicted ones.  It can be concluded that

\begin{enumerate}
\item The OC $R_c$ curve is well reproduced by the Gaussian curve.
\item The sharp $k$-space curve is remarkably different both from the
Gaussian and from the OC curve; then, Gaussian smoothing is definitely
preferred.
\item Remarkably, the sharp $k$-space curve is similar to the MS
curve.
\end{enumerate}

\begin{figure}
\plottwo{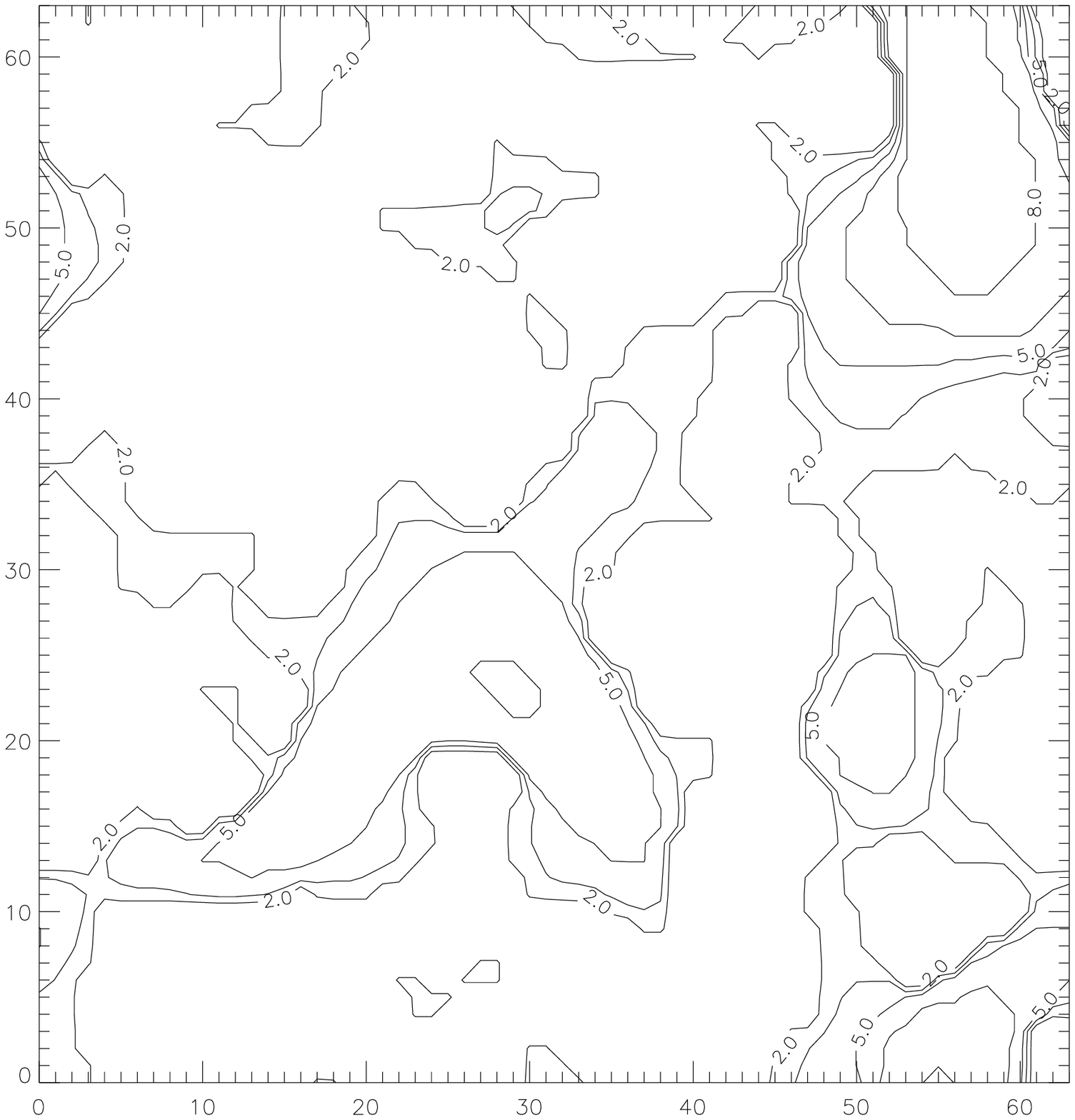}{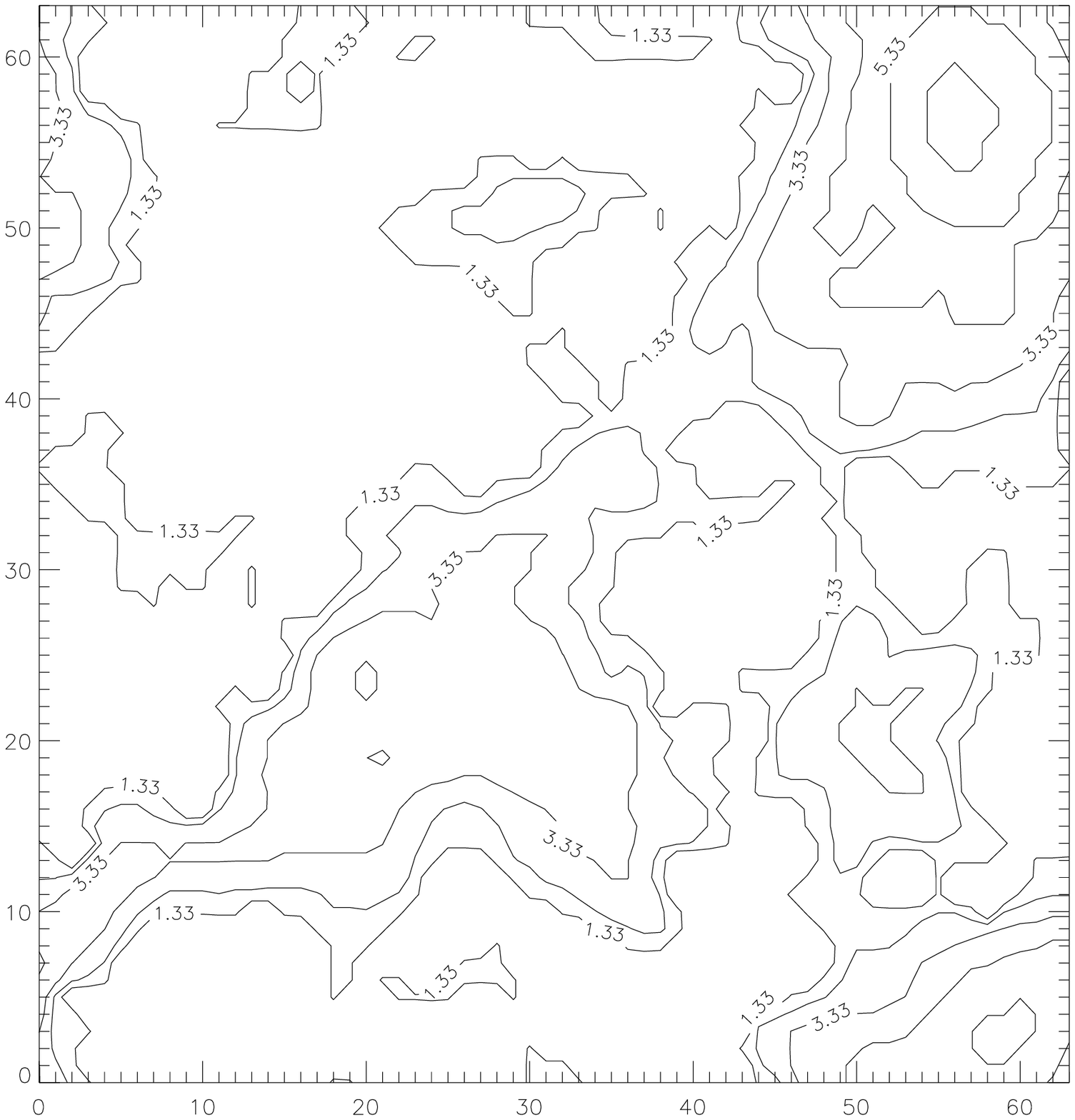}
\caption{Countour levels of the $R_c$ curve in 3D (evaluated in a
plane cut of the simulation box).  Left: ellipsoidal collapse. Right:
OC from the simulation.}
\end{figure}

It is possible to extend this analysis to 3D.  This has been done with
a test P$^3$M simulation with $64^3$ particles on a $64^3$ grid.  In
this case the OC regions are found by smoothing the vector field ${\bf
x}({\bf q})$, obtained from the output of the N-body simulation, in
the Lagrangian space, then calculating the Jacobian determinant
$J({\bf q})=\det(\partial x_a/\partial q_b)$ in each point; the points
with negative Jacobian determinant are in OC.  The curves $R_c({\bf
q})$ are then constructed.  Fig. 5 shows the contour level of two
$R_c$ curves, evaluated on a plane cut of the simulation box.  The
$R_c$ curves refer to the prediction of ellipsoidal collapse and to
the OC in the simulation.  Ellipsoidal collapse is able to reproduce
the OC regions in simulations fairly well.  On the other hand, linear
theory (not shown)turns out to be not bad in predicting the largest
structures, but misses completely the connected network, which gives
origin to small-scale objects (we are using Gaussian smoothing, in
which case the linear theory prediction is known to underestimate the
mass function).

\section{The mass function does not exist!}

As in the 1D case, the 3D $R_c$ curves tend to be, at large smoothing
radii, a collection of isolated and simply connected regions, the mass
of which saturates as $R$ decreases.  It is then possible to construct
an average growing curve for them, which can then be used to solve the
geometrical problem described above.  However, when the variance
becomes comparable to one, the collapsed regions become multiply
connected in a very messy way.  As a consequence, at variance with the
1D case, there is no obvious way to define the mass of structures; in
other words, it is not straightforward to construct an algorithm to
fragment the collapsed medium into ``structures''.  We are currently
trying to find reliable fragmentation algorithms; the goodness of a
fragmentation algorithm can be decided by seeing how well it
reproduces the structures found in some other way.  For instance, one
could construct a fragmentation algorithm able to reproduce the
friends-of-friends halos, but then another algorithm will be needed to
reproduce, e.g., the DENMAX halos, which are slightly different.
Moreover, every clump-finding algorithm is parametric (e.g., in the
friends-of-friends algorithm the linking length must be fixed), and
the results depend on the actual value used for the parameters.

There is no univocal way to find halos in simulations, and there is no
univocal way to fragment the collapsed medium into structures.  As a
consequence, there is no univocal way to define the mass function!
The only well-defined quantities are the $R_c$ curve shown in Figs. 3
and 5.

Luckily, this ``nihilist'' position is exaggerated from a practical
point of view: the largest structures are recovered more or less in a
similar way by different clump-finding algorithms, and the problem
comes as usual from the small-mass clumps, which come from the
multiply-connected medium (and from the left-overs of the largest
structures).  Anyway, this implicit uncertainty must be considered
when judging, for instance, the agreement of a mass function theory
with a friends-of-friends curve, as that shown in Fig. 2.

\section{Work in progress}

The dynamical mass function theory is currently been analyzed and
applied to many interesting problems.  The comparison to 3D
simulations will be soon performed with the large ($360^3$ points)
simulations already used by Governato et al. (1998).  

Once a reliable fragmentation algorithm is found, it can be used to
generate large catalogues of objects.  The semi-analytical predictions
can be extended to the spin and merging history of objects: the spin
is acquired from tidal torques during the mildly non-linear evolution
(Catelan \& Theuns 1996), while merging histories are found through
the excursion set approach (Lacey \& Cole 1993), even though the role
of the filter in this case must be clarified.  This can have some
definite advantages with respect to the standard N-body simulations,
as the semi-analytical calculations are much, much faster.  It can be
used, for instance, to simulate halos in deep pencil beams, or to find
galactic halos in Hubble volume-sized realizations.

\section{Why the hell does the Press \& Schechter work?}

None of the problems presented in this paper have been considered by
PS; nonetheless, their formula appears to work quite well.  Table 1
contains some of the elements which should be taken into account to
construct a fully rigorous theory for the mass function; it is
indicated how these element influence the large mass tail of the mass
function.  We have seen that non-linear (and non-spherical) dynamics
gives an increase of large mass objects because of tidal forces.  But
the PS is based on sharp $k$-space smoothing, while Gaussian smoothing
is preferred; this leads to a decrease of large-mass objects.  On the
other hand, OC neglects the matter infalling on the structure, which
is already in multi-stream regime; taking this into account increases
the mass of objects.  But OC mixes ``virialized'' clumps with
filaments; taking them out of the game would of course decrease the
mass of objects.  The deconvolution procedure described in Section 3
would generally decrease the number of large mass objects, because of
the width of the differential growing curve.  Add to this the rather
unpredictable value of the average $M(R)$ relation.

\begin{table}
\caption{Neglected elements in the PS}
\begin{center}
\begin{tabular}{cc}
Element & Effect \\
        & on large masses \\ 
\tableline
Dynamics & Increase \\ 
Gaussian filtering & Decrease \\
MS vs OC & Increase \\
Exclude filaments & Decrease \\
Deconvolution & Decrease \\
Average $M(R)$ & ??? \\
\end{tabular}
\end{center}
\end{table}

In practice, the PS numerical recipe is the simplest way to construct
a mass function from the Gaussian statistics of the initial density.
It is as if some kind of ``central limit theorem'' were in act: the PS
works because it neglects so many things that they all tend to
compensate each other! especially if some free parameter can be tuned
to obtain good fits to numerical simulations.

\acknowledgements

The author thanks Paolo Catelan, Alfonso Cavaliere, George Efstathiou,
Giuseppe Murante, Sergei Shandarin and Tom Theuns for many
discussions, and Fabio Governato for kindly providing the object
catalogues taken from his simulations.

\end{document}